\documentclass[useAMS,usenatbib]{mn2e}
\usepackage{graphicx,graphics,amsmath}

\begin{document}
\title[Flares, pulse peak broadening and QPOs in 4U 1901+03
]{Flares, Broadening of the Pulse frequency peak, and Quasi Periodic Oscillations in the Transient X-ray Pulsar 4U 1901+03}

\author[Marykutty James, Biswajit Paul, Jincy Devasia and Kavila Indulekha]{Marykutty James$^{1,2}$\thanks{E-mail: marykuttykjames@yahoo.co.in}, Biswajit Paul$^{2}$, Jincy Devasia$^{1,2}$ and Kavila Indulekha$^{1}$\\
$^{1}$School of Pure and Applied Physics, Mahatma Gandhi University, Kottayam-686560, Kerala, India\\
$^{2}$Raman Research Institute, Sadashivanagar, C. V. Raman Avenue, Bangalore 560080, India}

\maketitle

\begin{abstract}
After a long quiescence of three decades, the transient X-ray pulsar 4U 1901+03 became highly active in 2003 February. 
From the analysis of a large number of {\em Rossi X-ray Timing Explorer}/{\em Proportional Counter Array} (RXTE/PCA)
 observations of this source, we report here the detection of X-ray flares, a broadening of the pulse frequency feature
 and Quasi Periodic Oscillations (QPOs). The X-ray flares showed spectral changes, had a duration of 100 s - 300 s, and 
were more frequent and stronger during the peak of the outburst. In most of the observations during the outburst we also 
detected a broadening of the pulse frequency peak. We  have also found intensity dependent changes in the pulse profile at very short timescales.
  This reveals a coupling between the periodic and the low frequency aperiodic variabilities. In addition, near the end of the
 outburst we have detected a strong QPO feature centered at $\sim$0.135 Hz. The QPO feature is broad with a quality factor of
 3.3 and  with an rms value of $18.5\pm3.1\%$. Using the QPO frequency and the X-ray luminosity during the QPO detection period 
we estimated the magnetic field strength of the neutron star as 0.31$\times$10$^{12}$ G which is consistent with the
 value inferred earlier under the assumption of spin equilibrium.

\end{abstract}
\begin{keywords}binaries: general --- stars: individual
(4U 1901+03) --- X-rays: binaries --- X-rays:
individual (4U 1901+03) --- X-rays: stars
\end{keywords}

\section{INTRODUCTION}

Most of the accretion powered X-ray pulsars belong to the class of High Mass X-ray Binaries. The companion star may be a Be star or
 an OB supergiant. Most of the Be star sources are transients which are normally detected during outbursts. There are two kinds of
outbursts, normal outburst (Type-I) and giant outburst (Type-II). Be stars have circumstellar disks and when the neutron star 
passes through the disk, it produces Type-I outbursts.
 The Type-II outbursts are produced during the episodes of large mass outflow from the Be star. Transient X-ray pulsars are excellent
 candidates to investigate properties of the accretion process that are intensity dependent.

The recurrent  High Mass X-ray binary pulsar  4U 1901+03  was discovered by {\em Uhuru} and {\em Vela 5B} satellite in 1970-1971 
which was the  first detection of an outburst from this source \citep{Forman1976, P&T1984}. A second outburst in 2003 was reported by {\em Rossi X-ray Timing Explorer-All Sky Monitor} (RXTE-ASM). After the detection by ASM, detailed long term observations of this source was carried out with the {\em Proportional Counter Array} (RXTE-PCA).

RXTE observations revealed  pulsations of 2.763 s  and the pulse frequency variation suggested an orbital period of 22.58 days and eccentricity of about 0.035. No optical or IR counterpart has been found for this source \citep{Galloway2005}. 4U 1901+03 is included in the unusual class of X-ray binaries with wide orbits ($P_{orb}\geq20$ days) but low eccentricity. The source was also observed  in the hard X-ray band with the {\em International Gamma-Ray Astrophysics Laboratory} (INTEGRAL) between 2003 March 10  and  2003 April 13 \citep{Molkov2003}. \cite{Chen2008} carried out a pulse profile analysis of this source and found that the  pulse profile is luminosity dependent. A phase resolved spectroscopic study revealed a strong pulse phase dependence of the spectrum \citep{Lei2009}.

In the following sections we present a detailed timing analysis of RXTE/PCA archival data of 4U 1901+03 and report the discovery of several interesting features including flares, a broadening of the spin frequency peak, and a QPO at $\sim$0.135 Hz. 

\section{OBSERVATIONS AND DATA REDUCTION}

The HMXB pulsar 4U 1901+03 was observed with the RXTE in 2003 during an outburst after a long interval of quiescence. The RXTE/PCA observations started on 2003 February 10 and lasted for around 5 months. Though the RXTE/PCA has a large field of view of 1 square degree,  any contamination from nearby sources is negligible because there are no bright X-ray sources within 1$^\circ$  from this object. There were a total of 68 pointings during this time period and  data is available for  a total useful exposure of 398 ks. The PCA instrument of RXTE consists of five Proportional Counter Units (PCU) covering an energy range of 2-60 keV with an effective area of 6500 cm$^{2}.$ The ASM is sensitive to X-ray  photons between 1.5-12 keV. Table~\ref{pca-obs} gives a log of the observations. For our timing analysis we used data from all the PCA observations and during most of the observations PCU0 and PCU2 were operational. The data reduction was performed using the software package FTOOLS version 6.5. The background count rate  was simulated using the tool {\em pcabackest} and was subtracted from the total count rate. Appropriate  background models are used for the background simulation. Caldb version 1.0.1 and pcarsp version 11.7 were used for the creation of the energy response matrix of the PCA detections.

\begin {table}
\centering
 \caption{Log of PCA Observations}
~\\
\begin {tabular}{|c|c|c|c|}
 \hline
Year&Obs Ids&No. of &Total Durations\\
    &       &Pointings&(ks)\\
\hline
2003&P70068&5&111.8\\
    &P70096&63&286.5\\
    \hline
\end{tabular}
\label{pca-obs}
\end{table}

\begin{figure}
 \centering
 \includegraphics[width=6.5 cm, angle=-90]{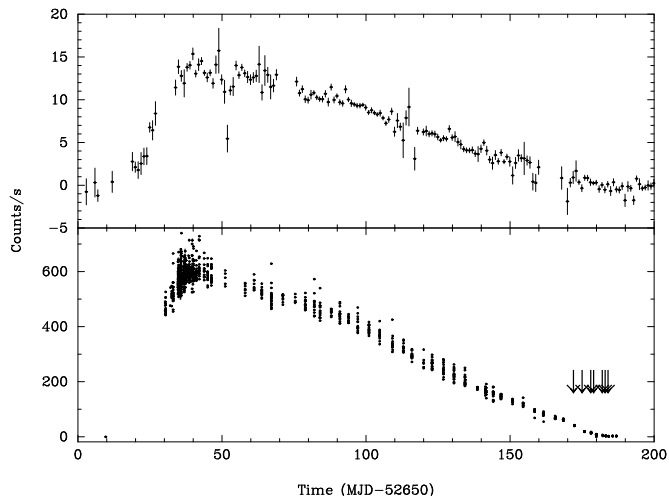}
 \caption{The top panel shows the one day averaged ASM light curve and the bottom panel shows 
the background subtracted PCU2 light curve of 4U 1901+03 with a bin size of 200 s. The arrow marks 
indicate the period of QPO detection.}
 \label{asm-pcu}
 \end{figure}

\section{ANALYSIS AND RESULTS}
\subsection{Light curve and Flares}
Many accretion powered pulsars show aperiodic variability, like flares, with timescales of a few seconds to a few hundred seconds. Here we are presenting results from the analysis of data from the RXTE/PCA long term observations of 4U 1901+03, in which we have detected a large number of X-ray flares.

One day  averaged  ASM light curve is shown in the top panel of Figure~\ref{asm-pcu} for a 250 day period around the outburst. The outburst started around MJD 52660 and reached a maximum count rate of 18 s$^{-1}$ in the ASM detectors (about 240 mCrab) \citep{Galloway2005} and then it declined linearly reaching  a count rate of about 1 counts/sec at MJD 52840. We extracted light curves from all the 68 PCA observations with a time resolution of 0.125 s. The entire background subtracted PCU2 light curve from all 68 observations is shown in the bottom panel of Figure~\ref{asm-pcu}. This 2-60 keV light curve is generated with a time resolution of 200 s. The arrow marks in Figure 1 indicate the periods 
of QPO detection described later.

\begin{figure}
 \centering
 \includegraphics[width=9.0 cm, angle=-90]{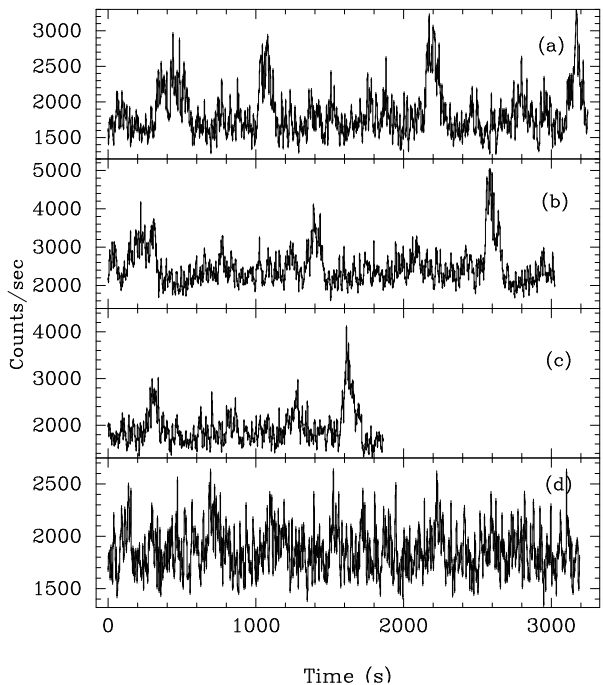}
 \caption{Some representative RXTE/PCA light curves of 4U 1901+03. The binsize used here equals the spin period of the neutron star. The start time of the light curves are (a)52685 15:34 (b)52685 21:57 (c)52689 13:08 (d)52686 18:25 MJD respectively}
 \label{flare-lc}
 \end{figure}

In most of the observations carried out during the outburst, 4U 1901+03 exhibited X-ray flares. 
The flares occurred randomly with different recurrence time between a few hundred to thousand seconds and flare duration of 100 
to 300 seconds. Pulsations are also clearly seen during the flares. Some of the flares had sharp rise and slow decay with 
multiple peak structure. In Figure~\ref{flare-lc} we have shown three light curves selected 
on the basis of intensity of flare  
activity and a fourth without flares but strong short term variability. The light curves are plotted with a bin size equal 
to the spin period of $\sim$2.763 s so that the variability associated with the pulsation is omitted. At the peak of the flare, the 
luminosity is higher by a factor of upto 2.5. The flares are found to be more frequent and stronger near the peak of the outburst 
indicating a relation with the mass accretion rate.
 
\begin{figure}
 \centering
 \includegraphics[width=7.5 cm, angle=-90]{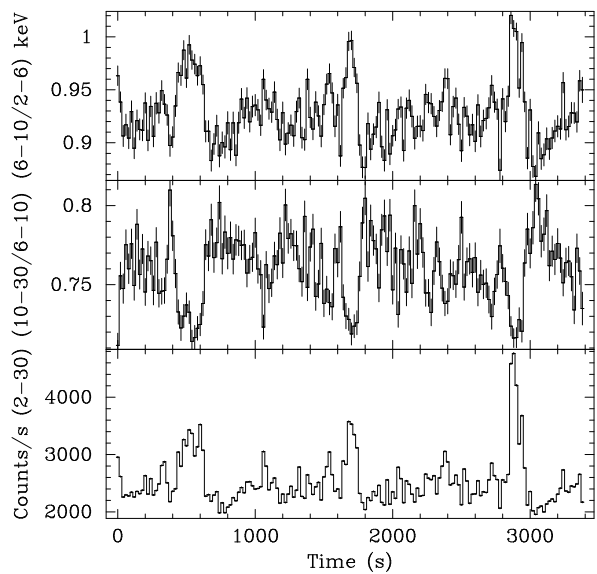}
 \caption{Background subtracted RXTE/PCA light curves (Figure 2b: 52685 21:57 MJD) and hardness ratios in two different energy band pairs are shown here.}
 \label{hardness-ratio}
 \end{figure}

We calculated the hardness ratios (soft hardness ratio: 6-10 keV / 2-6 keV and hard hardness ratio: 10-30 keV / 6-10 keV) of the light curves using the event mode data (Good Xenon mode) from the PCA. The hardness ratios of the data corresponding to panel (b) of Figure~\ref{flare-lc} are shown in Figure~\ref{hardness-ratio} along with the summed light curve in the bottom panel. During the flares, the soft hardness ratio is found to increase while the hard hardness ratio is seen to be decreasing. This implies that the flares are stronger in the medium energy band which we have also found in the analysis of the flare spectrum described later. A detailed hardness ratio analysis showed that all the flares have similar energy dependence.

\subsection{Power density spectrum}

Quasi Periodic Oscillations (QPOs) have been reported in many Accretion Powered Pulsars, both HMXBs and LMXBs (see James et al. 2010 for a complete list of the sources). In the accreting high magnetic field X-ray pulsars, the QPO  appears as a broad peak in the power density spectrum. We have carried out a timing analysis using data from all the RXTE-PCA observations to search for QPOs and other 
aperiodic features in this pulsar.

Standard1 mode data  of RXTE/PCA with a time resolution 0.125 s was used for the timing analysis. We extracted light curves from all the archival data and created Power Density Spectrum (PDS) using the FTOOL-{\em powspec}. The light curve was divided into stretches of length 1024 s and the power spectra obtained from 2-5 consecutive segments were averaged to produce a set of averaged PDS from which the expected white noise level was subtracted. The PDS were normalised such that normalised integral gives the squared rms fractional variability.

\begin{figure}
 \centering
 \includegraphics[width=9.0 cm, angle=-90]{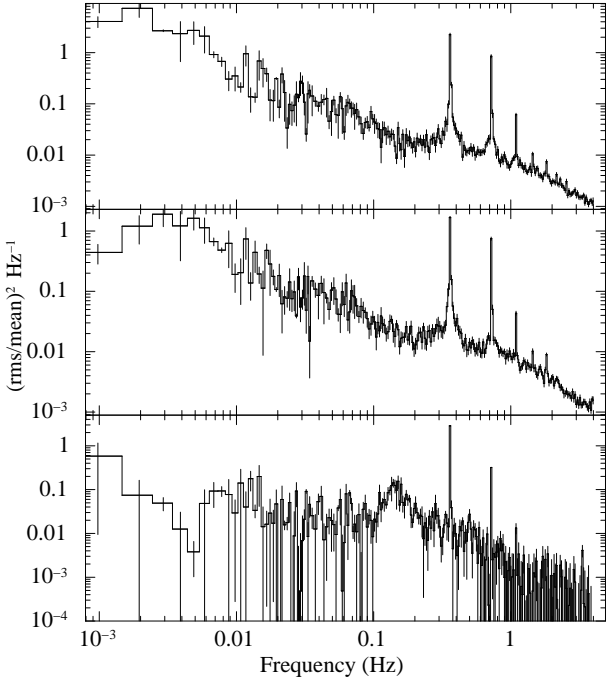}
\caption{Power density spectra with broadening (top and middle panel) and without broadening of the pulse peak
(bottom panel) are shown  in this figure.}
 \label{pds}
 \end{figure}

For most of the observation period, the PDS was a featureless continuum except for the pulse peak and its harmonics. 
The PDS reveals no QPO feature during the time of the outburst but in many of the observations we detected a broadening of the
 spin frequency peak.  Near the end of the outburst, the broadening of the spin frequency peak disappeared and a broad QPO 
feature at $\sim$0.135 Hz appeared in the PDS. The three PDS with and without flares are shown in Figure~\ref{pds} 
 (top panel: PDS for the period of flare; middle panel: non flare; bottom panel: QPO detection). 
 During the QPO detection we also detected a large pulsed fraction (upto 60\%)
and also a large rms (18\%) of the QPO. If these two features were to be from two different
soruces in the FOV, then even if the pulsating source was to be 100\% pulsed, the source
with QPOs is required to have a rms of 45\%. This is much larger than any QPO feature in
any type of X-ray binary. We, therefore, conclude that the QPO feature is indeed from the
X-ray pulsar. The PDS with the QPO 
was fitted with a model consisting of a Power law and Lorentzian for the continuum and a second Lorentzian for the
 QPO feature. The frequency bins corresponding to the pulse peak and its harmonics seen in Figure~\ref{pds} were ignored 
while fitting the continuum. Center frequency of the QPO feature was found to change from a value of $0.143\pm0.003$ 
 Hz to $0.130\pm0.006$ Hz in 3 days. The QPO feature has a width of about 0.04 Hz and a quality factor of about 3.3. 
The significance of the detection of the QPO was different on different days varying from  6 $\sigma$ to 7 $\sigma$.
 The background corrected rms value of the QPO feature is $18\pm3\%$. We calculated the energy dependence of the QPO rms using the event mode data (Good Xenon mode) of the PCA and found that the QPO feature is not detectable above 11 keV.

\begin{figure}
 \centering
 \includegraphics[width=9.0 cm, angle=-90]{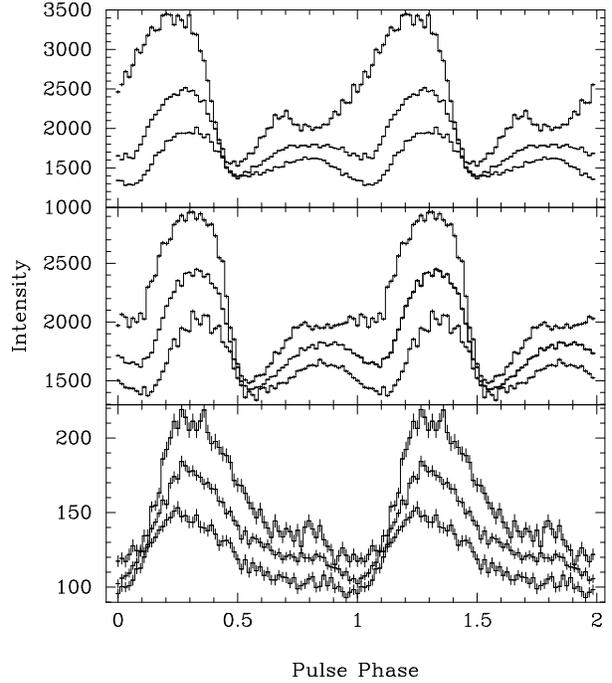}
 \caption{In each panel of this figure, pulse-profiles at three different intensity levels are shown.
 Top panel: when the light curve shows flares (Figure 2b); middle panel: when the light curve does not 
show flares (Figure 2d); bottom panel: when the light curves shows QPOs and no broadening of the pulse frequency peak (52825 18:09 MJD).}
 \label{pulse-pro}
 \end{figure}

A broadening of the X-ray pulse frequency peak has been seen in several other sources \citep{ Lazzati1997, Reig2006, Raichur&Paul2008, Jain2010} and is thought to be due to a coupling of the low frequency variations with the pulsation. To investigate this in detail, we have created intensity dependent pulse profiles. Light curve from RXTE-PCA observations were first created with bin size same as the pulse period. Using the X-ray intensity of each single pulse, separate intensity dependent time windows were created and these windows were then used to obtain intensity dependent pulse profiles of 4U 1901+03. The intensity dependent pulse profiles obtained from two light curves, with and without flares (panel b \& d in Figure~\ref{flare-lc}) are shown in the top two panels of Figure~\ref{pulse-pro}. 
It is to be noted that broadening of the pulse frequency peak is present in all the light curves shown in Figure~\ref{flare-lc}. In bottom panel of Figure~\ref{pulse-pro} we have shown the intensity selected pulse profiles during a period when the light curve shows a QPO feature but no broadening of the X-ray pulse frequency peak. Note that the three PDS shown in Figure~\ref{pds} and Figure~\ref{pulse-pro} are from the same set of light curves (top: flare, middle: non-flare, and bottom: QPO).

In the top two panels of Figure~\ref{pulse-pro} one can clearly see that at low intensity, the pulse profile has two unequal peaks. At higher intensity, both the peaks become stronger, the dip before the main peak disappears while the dip after the main peak becomes more clear. The intensity dependence of the pulse profile appears to be quite similar  for the flaring and non-flaring episodes. From the bottom panel of Figure~\ref{pulse-pro} we can see that the episode that does not show a pulse peak broadening also does not have any significant intensity 
dependence of the pulse profile.

%   \begin{figure}
%   \centering
%   \includegraphics[width=6.0 cm, angle=-90]{fig5.ps}
%   \caption{Power density spectra.}
%   \label{qpo}
%   \end{figure}

\subsection{Spectral Analysis}

\begin{figure}
 \centering
  \includegraphics[width=6.5 cm, angle=-90]{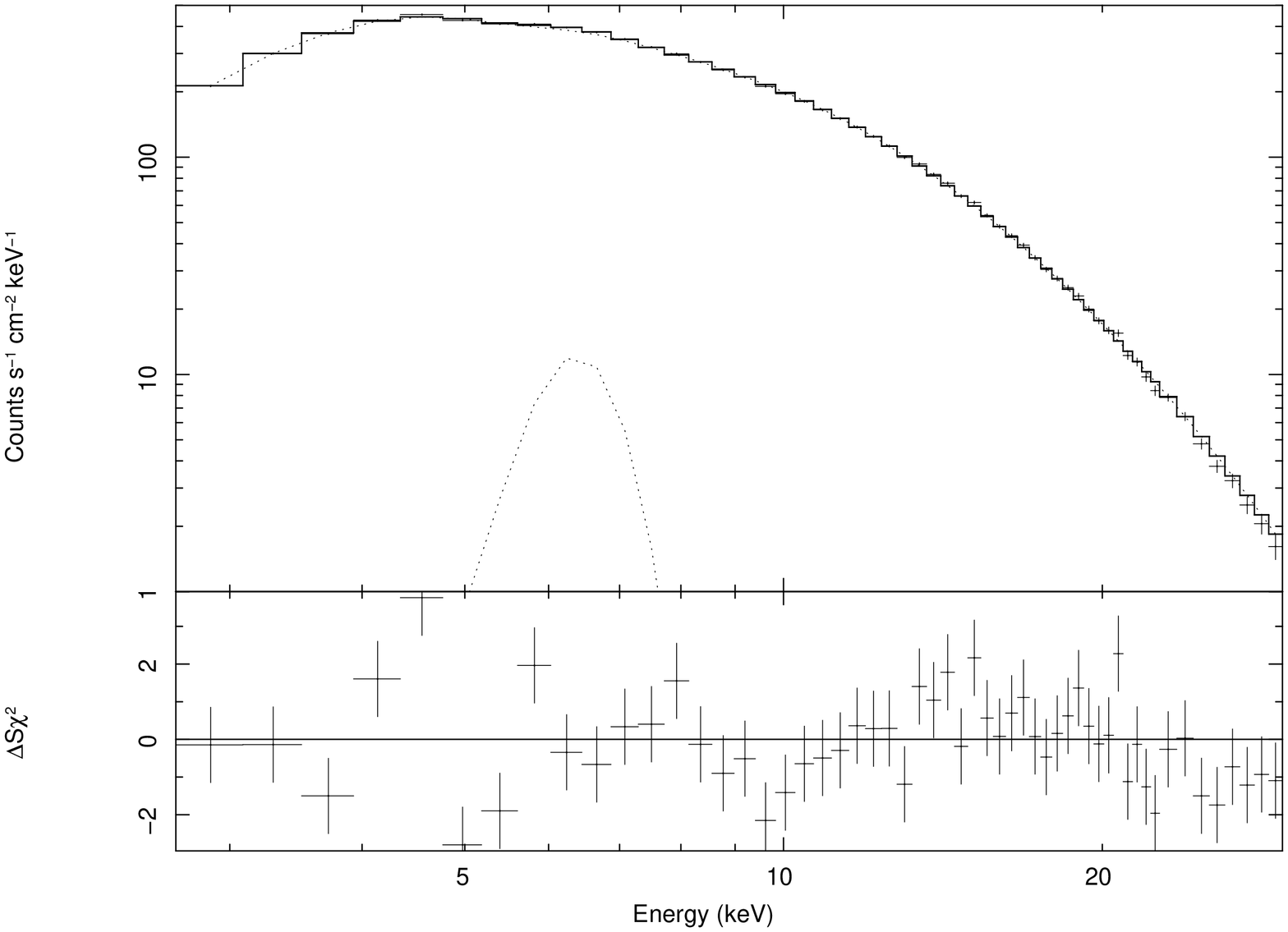} \includegraphics[width=6.5 cm, angle=-90]{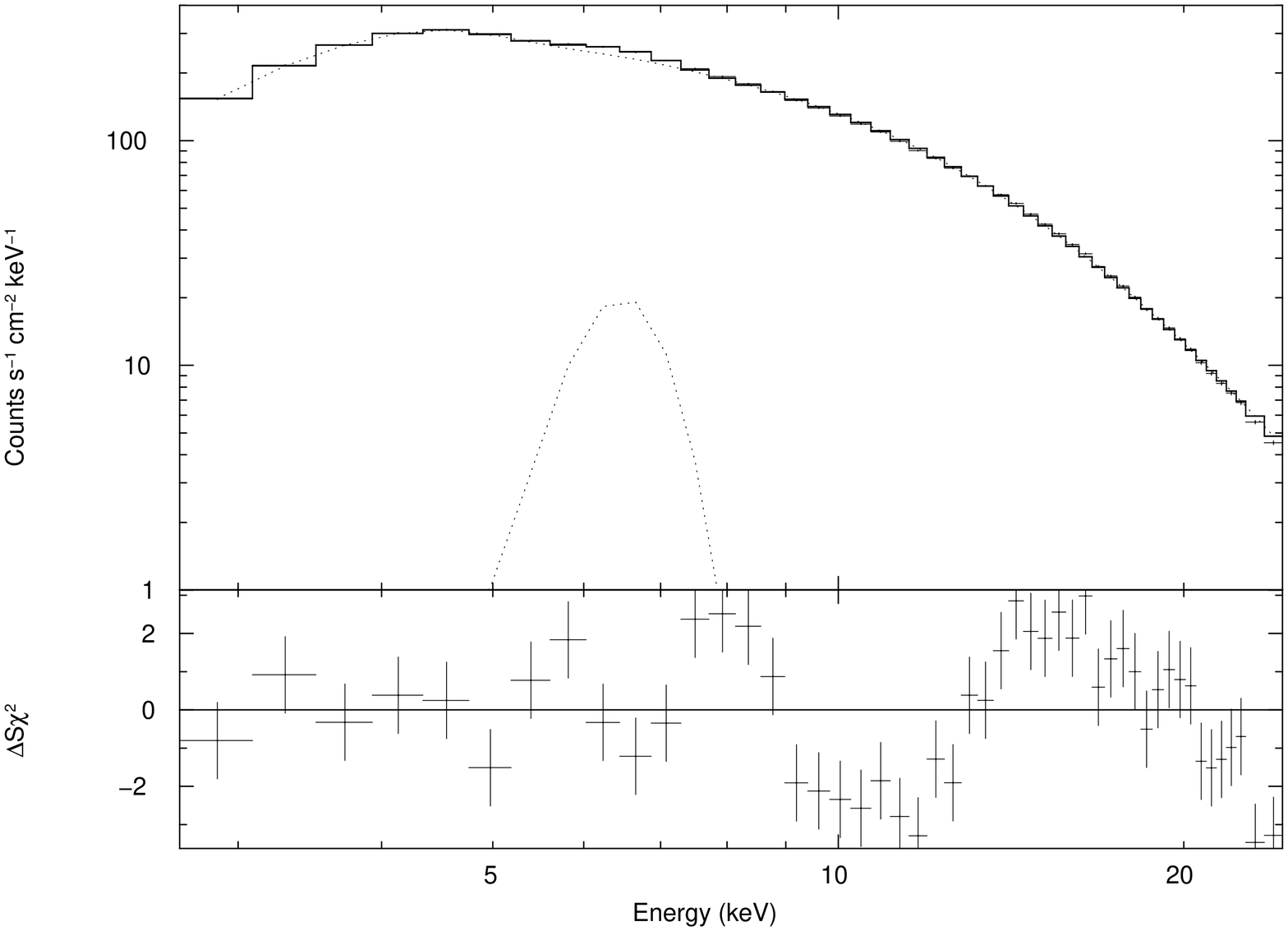}
 \includegraphics[width=6.5 cm, angle=-90]{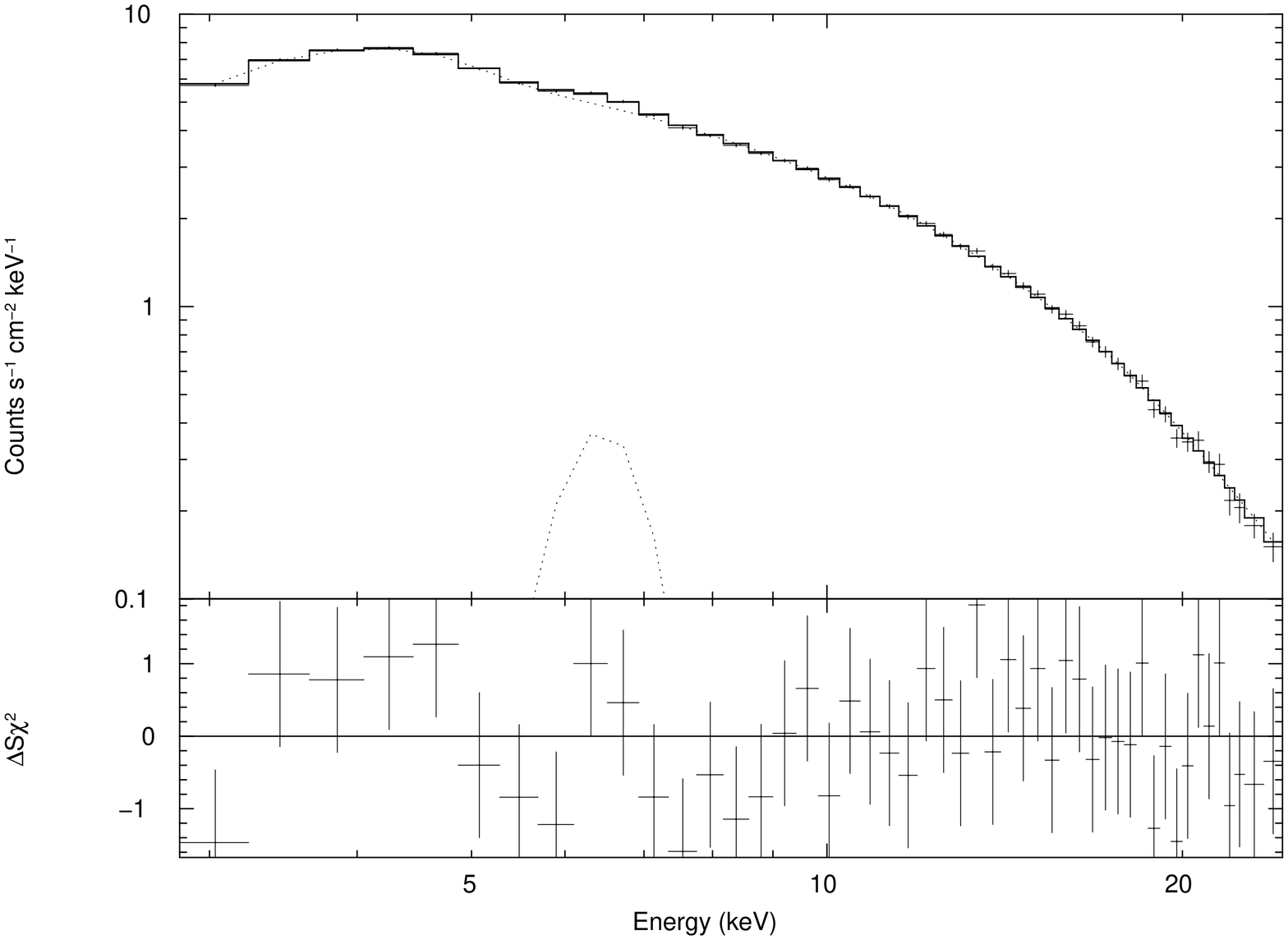}
 \caption{The  2.5-25 keV spectra of 4U 1901+03 during X-ray flares (top), non-flares (middle) and
during QPO detection (bottom) are shown here along with the respective best fit models consisting of absorption, power law, high-energy cutoff and gaussian components.
 The lower panel in each figure shows the residual to the best fit models.
}
 \label{spectrum}
 \end{figure}

We have done a spectral analysis of the source during the QPO detection, flaring and non flaring state to investigate if
 there are any differences. Standard2 data of the PCA were used for extraction of the energy spectrum in 2.5-25 keV 
energy band. All the three spectra were fitted with a model consisting of an absorption component, a power law with a high-energy
 cutoff and a gaussian emission line. A systematic error of $0.5\%$ was added to the first and second spectra to account for the 
calibration uncertainties while the non-flaring spectrum required a systematic error of $1\%$. The spectra, along with the 
respective best-fit models and the residuals are shown in Figure~\ref{spectrum} (top panel: spectrum for the period of flare; middle panel: non flare; bottom panel: QPO 
detection). The best fitted spectral parameters are given 
in Table~\ref{spec-para} along with the reduced $\chi^2_r$ values. The spectrum taken when QPOs are present and 
broadening of spin frequency peak is absent has a larger photon index and large value of the high energy cut-off compared
 to the spectrum when QPOs are absent and spin frequency peak shows a broadening.
%The QPO spectrum gives a $\chi^2_r$ of  0.9 with 44 d.o.f while the flaring and 
non flaring spectra gives $\chi^2_r$ of 1.3 and 1.6 respectively with  42 d.o.f.
If we take a ratio of the flaring and the non flaring spectrum, 
the medium energy band of 6-10 keV shows the maximum difference. This is in
 accordance with the hardness ratio changes during the flares (Figure~\ref{hardness-ratio}). 
The flux measured in the 2-20 keV band during the time of the QPO detection is
 about 1.3$\times$10$^{-10}$ erg cm$^{-2}$sec$^{-1}$. The flux measured 
in the 2-20 keV band using the spectra shown in Figure~\ref{spectrum} during the 
time of flare and non flare time are  7.7$\times$10$^{-9}$ erg cm$^{-2}$sec$^{-1}$ 
and 5.2$\times$10$^{-9}$ erg cm$^{-2}$sec$^{-1}$ respectively. 

{
\begin {table}
 \caption{Best fit Spectral Parameters of 4U 1901+03}
~\\
\begin {tabular}{|c|c|c|c|}
 \hline
 \hline

Parameter&Flare&Nonflare&QPO \\
 \hline
$N_{H}$ (10$^{22}$ cm$^{-2}$)&3.96$\pm$0.08&3.03$\pm$0.62&0.283$\pm$0.09\\
$\Gamma$& 0.98$\pm$0.15&0.79$\pm$0.06&1.55$\pm$0.008\\
norm& 0.46$\pm$0.087&0.24$\pm$0.03&0.016$\pm$0.00045\\
$E_{c}$ &6.29$\pm$0.52&4.56$\pm$0.33&17.07\\
$E_{f}$&9.46$\pm$0.71&9.20$\pm$0.45& 19.46\\
$E_G$(keV)&6.57$\pm$0.28&6.56$\pm$0.09&6.42$\pm$0.069\\
GN&0.0031$\pm$0.0006&0.0049$\pm$0.0009&0.000086$\pm$0.000013\\
EQW(keV)&0.04&0.12&0.1\\
$\chi^2_{r}$&1.37&1.62&0.9\\
\hline
% $N_{H}$ and $\Gamma$ are the hydrogen column density and photon index.\\
% $E_{c}$, $E_{f}$ and $E_{G}$ are the cut off energy, folding energy and Gaussian line.\\
\end{tabular}
\label{spec-para}
\end{table}

\section{DISCUSSION}

Detailed timing and spectral studies of 4U 1901+03 using this set of RXTE-PCA observations have been reported earlier. 
This includes timing and spectral evolution studies over the duration of the outburst, pulse frequency evolution 
(Galloway et al. 2005), energy and luminosity dependence of the pulse profile (Chen et al. 2008) and pulse phase dependence 
of the energy spectrum (Lei et al. 2009). In the previous section we have reported some additional temporal studies, 
including the detection of an intensity dependence of the pulse profile at very short time scales; probably for the first time in any short period pulsar.

\subsection{Flares}

X-ray Flares of different duration, occurrence and flux enhancement have been reported in many accretion powered pulsars like 
GRO J1744-28 \citep{Finger1996}, 
SMC X-1 \citep{Moon2003c}, 
EXO 2030+375 \citep{Parmar1989}, 
LMC X-4 \citep{Moon2001b, Moon2003d}, 
4U 1907+09 \citep{intzand1998, Mukherjee2001}, 
Vela X-1 \citep{Kreykenbohm2008},
SWIFT J1626.6-5156 \citep{Reig2008}, 
Her X-1 \citep{Moon2001a} and
4U 1626-67 \citep{McClintock1980}. 

The extremely large flux enhancements seen in GRO J1744-28 and the large flares in SMC X-1 are probably due 
to accretion disk instability similar to the bursts in the rapid burster. The large flares in EXO 2030+375 and
 LMC X-4 are similar, and can be caused by Rayleigh-Taylor Instability at the magnetosphere or variable mass 
accretion from a clumpy stellar wind. The flares from the above sources lack any associated spectral change.
 The fast rise and slow decay as seen in some of the flares in 4U 1901+03 are similar to those in the flares 
in EXO 2030+375, though the latter were of much longer duration. The 1000 s flares from 4U 1626-67 are caused 
by carbon burning under the neutron star surface. The flares in 4U 1901+03 show a spectral hardening, while
 a spectral softening is expected if these flares had been of thermonuclear origin. Somewhat smaller amplitude
 flares are seen in 4U 1907+09, Vela X-1 and SWIFT J1626.6-5156 which are most likely caused by variable wind accretion.
 Overall, the flares in 4U 1901+03 closely resemble the flares in SWIFT J1626.6-5156 in their duration, flux enhancement, and spectral variation. It is therefore quite likely that the flares in 4U 1901+03 are also caused by variable accretion, either due to inhomogeneities in the accretion disk or small scale clumping in the stellar wind as is seen in some Supergiant Fast X-ray Transients (Sidoli et al. 2010 and references therein).

\subsection{ Broadening of the pulse frequency peak and intensity dependence of pulse profiles}

A significant  broadening of the pulse frequency was discovered in sources like Cen X--3, 4U 1145--62 and Vela X-1 using EXOSAT
 observations. It was proposed that the coupling between the periodic and aperiodic variability appears as broadening in the
 wings of the spin frequency peak \citep{Lazzati1997}. Multiple couplings were discovered in Her X-1 from RXTE-PCA observations
 \citep{Moon2001a}. Recently, using {\it RXTE}-PCA observations of several sources, a broadening of the pulse frequency peak 
has been shown: Cen X-3 \citep{Raichur&Paul2008}, 4U 1626--67 \citep{Jain2010} and KS 1947+30 \citep{james2010}. For most of 
the outburst period of 4U 1901+03, we have found a broadening of the pulse frequency peak that we attribute to a
 change in the pulse shape at a short time scale of a few tens to a few hundreds of seconds. 
Interestingly, the  broadening of the pulse frequency peak and the pulse shape change are absent in the 
low state when the QPOs appear.

Over the entire outburst period, 4U 1901+03 showed a variety of pulse profiles \citep{Galloway2005}. 
It is interesting to note that some of the pulse profile changes reported in Galloway et al. (2005)
 at different intensity levels are similar to the profiles shown here, though the former were obtained at 
different overall intensity level with time difference of several days/weeks and the latter are acquired 
within a timespan of less than one hour. This indicates that the pulsar beaming pattern changes from pulse to pulse, 
depending on the instantaneous mass accretion rate.

\subsection{Quasi Periodic Oscillations}

Quasi Periodic Oscillations from Accretion Powered Pulsars have thus far been reported from 17 sources including the recent
 discovery from the transient pulsar KS 1947+300 (James et al. 2010). The first discovery of a QPO in an accretion powered pulsar was made in 1989, in the source EXO 2030+375 \citep{Angelini1989}. All types of accretion powered pulsars, i.e persistent and transient LMXB $\&$ HMXB pulsars have shown low frequency QPO features. 4U 1901+03 is the tenth high magnetic field transient pulsar in which QPOs have been detected.

In accretion powered pulsars, QPOs seem to occur more in transient sources compared to the persistent ones. In most of these cases, the QPO shows a transient nature. In KS 1947+300 and 4U 1901+03, the QPOs are detected during the end of the outburst only while a persistent QPO feature was seen in the transient pulsar XTE J1858+034 \citep{Paul&Rao1998}. Persistent pulsars like Cen X-3 do not show QPOs all the time \citep{Raichur&Paul2008} except for 4U 1626+67 \citep{Kaur2008} in which QPOs are detected in most of the observations.

There are many models that have been proposed to explain the Quasi Periodic Oscillations in X-ray binaries.
 Beat Frequency Model is generally considered as the most plausible model for QPOs in X-ray Pulsars.
 In this model, blobs of matter entrained in the neutron star magnetic field which orbit with the Keplerian frequency of
 the inner edge of the accretion disk are accreted at a rate that is modulated by the rotating magnetic field \citep{Alpar&Shaham1985}.
 It produces an aperiodic variability at the beat frequency between the pulsar spin frequency and the 
 Keplerian frequency; $\nu_{QPO}= \nu_{k} -\nu_{s}$. Another possible model is the Keplerian Frequency Model in 
which the accretion disk contains structures that persists for a few cycles around the neutron star and modulate 
the X-ray flux by obscuration. I.e $\nu_{QPO}= \nu_{k}$ \citep{Klis1987}.

In the present source and in some other sources like Cen X-3, 4U 0115+63, 4U 1626-67, Her X--1, LMC X-4, V 0332+52, SMC X--1
 and KS 1947+300, the pulsar frequency is higher than the QPO frequency hence the Keplerian frequency model is inapplicable. 
 It is expected that centrifugal forces inhibit mass accretion if the Keplerian frequency at the inner disk radius is less 
than the spin frequency of the pulsar. The X-ray flux is proportional to the mass accretion rate, and the Keplerian frequency at the inner edge 
of the accretion disk is related to the X-ray flux \citep{Finger1998}. So in both KFM and BFM we expect a positive correlation 
between QPO frequency and the luminosity which was however, not seen in two of the sources V 0332+53 \citep{Lu2005} and
 GRO J1744-28 \citep{Zhang1996}}. So neither of these models are applicable in these sources. But in sources like A0535+262, EX0 2030+375 and XTE J1858+034 \citep{Mukherjee2006} the QPO frequency depends on the X-ray intensity.

In the 2.7 s pulsar  4U 1901+03, the QPO is detected at $\sim$0.135 Hz  during decline phase of the 2003 outburst. The same kind of QPO feature was also discovered in KS 1947+300, during its 2001 outburst. KS 1947+300 source has the smallest QPO frequency among all the transient pulsars \citep{james2010}. In 4U 1901+03, the QPO frequency is less than the spin frequency of the pulsar, so it can be explained by the Beat Frequency Model. Radius of the QPO production region can be estimated as

\begin{equation}
R_{qpo} = \left({GM}\over{4 \pi^{2} \nu^{2}_{k}}\right)^{1/3}
\end{equation}

For the Beat Frequency Model, the Keplerian frequency is $\nu_{k}  \sim 0.5$ Hz. 
Assuming a neutron star mass of 1.4 $M_{\odot}$, the inner radius of the accretion disk is obtained as $2.6\times 10^{3}$ km.

The 2-20 keV X-ray flux during the time of QPO detection is 1.3$\times$10$^{-10}$ erg cm$^{-2}$ sec $^{-1}$.
 The total X-ray luminosity for the  source is calculated to be about $1.6\times10^{36}$ erg s$^{-1}$ at 10 kpc. The magnetospheric radius of a neutron star can be expressed in terms of the luminosity and magnetic moment as \citep{Frank&Raine2002},

\begin{equation}
R_{m}= 3\times10^{8}L^{-2/7}_{37}\mu^{4/7}_{30}  ~\\ cm
\end{equation}

where $L_{37}$ is the X-ray luminosity in units of 10$^{37}$ erg s$^{-1}$ and $\mu_{30}$ is the
 magnetic moment in units of 10$^{30}$ cm$^{3}$  Gauss. Equating $R_{m}$  with $R_{qpo}$ we estimate 
the magnetic field as 0.31$\times$10$^{12}$ G. This is consistent  with the value inferred  by Galloway et al. (2005) by 
estimating magnetic field strength from spin equilibrium (0.3$\times$10$^{12}$ G). This is also within the bounds for 
the magnetic field derived from  the upper and lower flux levels within which the source behaved as an X-ray pulsar. 
In the Beat Frequency Model, the expected QPO frequency range, for source luminosity between 
5.8 $\times$ 10$^{37}$ erg s$^{-1}$ at the peak of the outburst and 0.16$\times$10$^{37}$ erg s$^{-1}$ when the QPOs are detected, is 2.05 to 0.135 Hz. But only a 0.135 Hz QPO is detected in this source.

 \section{Acknowledgements}
We thank an anonymous referee for suggestions that helped us to improve the manuscript. This research has made use of data obtained through the High Energy Astrophysics Science Archive Research Center Online Service, provided by the NASA/Goddard Space Flight Center.

%............................

\def\etal{{\it et~al.\ }}
\def\apj{{Astroph.\@ J.\ }}
\def\apjl{{Astroph.\@ J. \@ Lett. }}
\def\araa{{Ann. \@ Rev. \@ Astron. \@ Astroph.\ }}
\def\mn{{Mon.\@ Not.\@ Roy.\@ Ast.\@ Soc.\ }}
\def\aap{{Astron.\@ Astrophys.\ }}
\def\aj{{Astron.\@ J.\ }}
\def\prl{{Phys.\@ Rev.\@ Lett.\ }}
\def\pd{{Phys.\@ Rev.\@ D\ }}
\def\nucp{{Nucl.\@ Phys.\ }}
\def\nat{{Nature\ }}
\def\sci{{Science\ }}
\def\plb {{Phys.\@ Lett.\@ B\ }}
\def \jetpl {JETP Lett.\ }

\end{document}